\def\BE{\begin{equation}}
\def\EE#1{\label{#1}\end{equation}}
\def\be{\begin{align}}
\def\ba{\begin{align*}}
\def\se#1{\begin{subequations}\label{#1}
\renewcommand{\theequation}{\theparentequation.\arabic{equation}}}
\def\rf#1{(\ref{#1})}
\def\xf#1{Fig.~\ref{#1}}
\def\H{{\cal H}}
\def\R{{\mathbb R}}
\def\T{{\mathbb T}}
\def\I{{\rm i}}
\def\e{{\rm e}}
\def\d{{\rm d}}
\def\L{\boldsymbol{\mathfrak{L}}}
\def\Ae{\mathfrak{A}}
\def\A{\boldsymbol{\mathfrak{A}}}
\def\sAe{{^{\rm s}\mathfrak{A}}}
\def\sA{{^{\rm s}\boldsymbol{\mathfrak{A}}}}
\def\De{\mathfrak{D}}
\def\D{\boldsymbol{\mathfrak{D}}}
\def\sD{{^{\rm s}\!D}}
\def\mD{{^{\rm s}\bf D}}

\def\LA{\left\langle}
\def\RA{\right\rangle}
\def\bF{{\bm\Phi}}
\def\bo{{\bm\omega}}
\def\wb{\widehat{\bf b}}
\def\wl{\widehat\lambda}
\documentclass[superscriptaddress,pre,reprint]{revtex4-1}
\usepackage{amsmath,amssymb,graphicx,bm}
\begin{document}
\title{Pointwise vanishing velocity helicity of a flow
does not preclude\\magnetic field generation}

\author{Alexander \surname{Andrievsky}}
\affiliation{Institute of Earthquake Prediction Theory and
Mathematical Geophysics, Russian Ac. Sci.,
84/32 Profsoyuznaya St, 117997 Moscow, Russian Federation}

\author{Roman \surname{Chertovskih}}
\affiliation{Research Center for Systems and Technologies (SYSTEC), Faculty of
Engineering, University of Porto,
Rua Dr.~Roberto Frias, s/n, 4200-465, Porto, Portugal}

\author{Vladislav \surname{Zheligovsky}}
\affiliation{Institute of Earthquake Prediction Theory and
Mathematical Geophysics, Russian Ac. Sci.,
84/32 Profsoyuznaya St, 117997 Moscow, Russian Federation}

\begin{abstract}
Pointwise zero velocity helicity density is shown not to prevent steady flows
from acting as kinematic dynamos. We present numerical evidences that such
flows can generate both small-scale magnetic fields as well as, by the magnetic
$\alpha$-effect or negative eddy diffusivity mechanisms, large-scale ones.
The flows are constructed as
curls of analytically defined space-periodic steady solenoidal flows, whose
vorticity helicity (i.e., kinetic helicity) density is everywhere zero.
\end{abstract}
\maketitle

\section{Introduction}\label{intr}

By a general mathematical definition, the helicity $\H_{\bf f}$ of a solenoidal
zero-mean field ${\bf f(x)}=\nabla\times\bF({\bf x})$ is the spatial integral of
the scalar product of the field and its vector potential $\bF$. When both
fields are space-periodic, the cell periodicity being the cube $\T=[0,2\pi]^3$,
\BE\H_{\bf f}=\int_\T{\bf f(x)}\cdot\bF({\bf x})\,\d{\bf x}.\EE{hel}
This quantity characterizes the knottedness of the field lines
of the field $\bf f$ \cite{MR}. Thus, helicities of different solenoidal
flow-related fields (for instance, the flow $\bf v$ itself and the vorticity
$\bo=\nabla\times\bf v$) constitute a set of parameters measuring the flow
complexity.

The magnetic $\alpha$-effect is supposed to play an important role in
generation of cosmic and planetary magnetic fields. It is a manifestation of
the interaction of fluctuating small-scale components of the field and velocity
of the generating flow \cite{P55}. Heuristically, a generating flow $\bf v$
is likely to have an intricate spatial structure featuring considerable
knottedness of lines of the flow-related fields (implying, by virtue
of the magnetic induction equation, that the magnetic field also has
a nontrivial small-scale structure), the respective helicities not vanishing.

In the dynamo studies, a prominent quantity is the {\it vorticity helicity}
$\H_\bo$ (often referred to as {\it kinetic helicity}). It is a hydrodynamic
invariant \cite{Mor,M69} of ideal fluid flow, constraining the topology
of vorticity lines \cite{MR}. The latter are a classical object described
by the Helmholtz theorems. The discovery \cite{MP}, that the helicity spectrum
\BE H_{\bf m}({\bf v})=\overline{\widehat{\bf v}}_{\bf m}\cdot
(\I{\bf m}\times\widehat{\bf v}_{\bf m})\EE{hsd}
(a pedantic note: in line with the general definition of helicity of vector
fields, it is logical to call it {\it the vorticity helicity spectrum}),
where $\widehat{\bf v}_{\bf m}$ are the Fourier coefficients of the flow,
$${\bf v(x)}=\sum_{\bf m}\widehat{\bf v}_{\bf m}\,\e^{\I\bf m\cdot x},$$
is crucial for the presence of the $\alpha$-effect in the limit of small
magnetic Reynolds numbers, has triggered many studies trying to establish
links between $\H_\bo$ and the flow dynamo properties (see also
\cite{SP}). However, the ability of a flow to generate magnetic field
does not require its vorticity helicity density to be non-zero in the physical
(the Christopherson flow is a counterexample
\cite{ZG}) or Fourier spaces (see \cite{GFP} for an example of a generating
flow with a zero helicity spectrum). Recently, a large variety of steady
solenoidal flows with a pointwise zero vorticity helicity and
zero helicity spectrum were shown \cite{RCZ} to act as kinematic magnetic
dynamos generating small-scale magnetic fields and, by the mechanisms
of the magnetic $\alpha$-effect and eddy (``turbulent'') diffusivity,
large-scale ones.

However, the knottedness of fluid particle trajectories is directly
characterized by the {\it velocity helicity} $\H_{\bf v}$ of the flow, rather
than by the vorticity helicity $\H_\bo$, and thus the quantity $\H_{\bf v}$
may be closer related to the dynamo properties of flows
(for instance, due to the frozenness of magnetic field in the ideal
magnetohydrodynamics, see~\cite{Mb}).

For turbulent flows, in the low conductivity limit the $\alpha$-effect is
proportional to the velocity helicity $\H_{\bf v}$ \cite{KR} (see also \cite{Mo}).
Generation of large-scale fields by the $\alpha$-effect was considered in
\cite{RBr} under assumptions in the spirit of the second-order
correlation approximation, and the results were supposed to be reliable
for small magnetic Reynolds numbers. The authors concluded that while for
flows varying rapidly in time the vorticity helicity $\H_\bo$ is the quantity crucial
for the presence of the $\alpha$-effect, for steady ones the velocity helicity
$\H_{\bf v}$ is crucial. Motivated by this statement, the present paper is devoted
to a study of this conjecture by considering small- and large-scale dynamos
for steady flows, whose velocity helicity density vanishes everywhere in space.

A comment is in order: The potential $\bF({\bf x})$ of a field
${\bf f(x)}=\nabla\times\bF({\bf x})$ is defined up to an arbitrary gradient.
While the helicity \rf{hel} is gauge-independent \cite{MR}, the helicity
density ${\bf f(x)}\cdot\bF({\bf x})$ does depend on the gauge; we consider
the scalar product of the field and its {\it solenoidal zero-mean}
space-periodic vector potential
$\bF({\bf x})=-\nabla^{-2}(\nabla\times{\bf f(x)})$, where
$\nabla^{-2}$ denotes the inverse Laplace operator.

Six families of steady solenoidal zero-mean flows $\bf w$,
\hbox{$2\pi$-periodic} in each Cartesian coordinate, that have zero vorticity
helicity, ${\bf w\cdot(\nabla\times w)}=0$, were constructed in~\cite{RCZ}.
For such a field $\bf w$, obviously, the flow
\BE\bf v=\beta\nabla\times w\EE{vw}
has a zero velocity helicity $\H_{\bf v}$; here $\beta>0$ is a normalization
factor such that the r.m.s.~flow velocity of \rf{vw} is 1. To construct
numerical examples, we use sample flows \rf{vf} for vector potentials $\bf w$
from families V$_1$, V$_2$, L \rf{wf} (see section \ref{al}) and flows \rf{cos}
for vector potentials $\bf w$ from family C \rf{wc} (see section \ref{ed}).
Since the helicity spectrum \rf{hsd}
of fields ${\bf w(x)}=\sum_{\bf m}\widehat{\bf w}_{\bf m}\,\e^{\I\bf m\cdot x}$
from families C, V$_1$ and V$_2$ vanishes \cite{RCZ}, the helicity
spectrum of the respective flow \rf{vw} is zero as well: for any $\bf m$,
$$H_{\bf m}({\bf v})=\overline{\I\bf m\times\widehat{w}_m}\cdot
(\I{\bf m}\times(\I{\bf m}\times\widehat{\bf w}_{\bf m}))=|{\bf m}|^2
H_{\bf m}({\bf w})=0.$$
For a sample flow \rf{vw} used here for computations,
where the potential $\bf w$ belongs to family L, see \xf{hs}. Family P flows
were defined {\it ibid.} as poloidal flows whose scalar vector potential
satisfies a certain partial differential equation. We do not
consider potentials $\bf w$ from family P, since the curl of any poloidal
field is toroidal and hence the respective flow \rf{vw} is planar, and
by the Zeldovich \cite{Zel} antidynamo theorem such flows are incapable
of the dynamo action.

\begin{figure}[t]
\centerline{\includegraphics[height=1.5in]{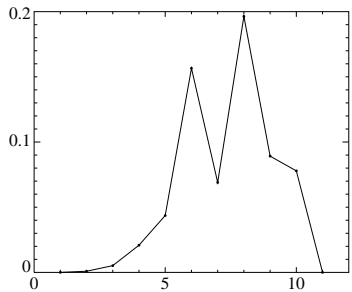}}

\caption{Helicity spectrum seminorms $\sum_{M-1<|{\bf m}|\le M}|H_{\bf m}({\bf v})|$
(vertical axis) vs $M$ (horizontal axis) for a sample flow \rf{vw} constructed
for a potential $\bf w$ from family L \cite{RCZ} (see the definition in section
\ref{al}.}
\label{hs}\end{figure}

The paper is organized as follows. In the next section we briefly review
the standard multiscale expansion yielding expressions for the tensors
of the magnetic \hbox{$\alpha$-effect} and eddy diffusivity in the limit of high scale
separation. In sections \ref{al} and \ref{ed} we present results of computation
of the dominant growth rates of large-scale harmonically amplitude-modulated
magnetic modes generated by the action of the $\alpha$-effect and negative
eddy diffusivity, respectively. In both cases, the growth rates may experience
a singular behavior in the vicinity of the critical molecular diffusivity
for the onset of the small-scale dynamo action. In the presence of negative
eddy diffusivity, this phenomenon is well-known (see sec.~3.7 in \cite{VZ}).
In the presence of the $\alpha$-effect, it was first noticed and qualitatively
explained in \cite{RCZ}. In the present computations we observe, that
at the critical molecular diffusivity the magnetic induction operator has
a Jordan normal form cell of size 2 associated with the zero eigenvalue.
Based on this observation, in section \ref{exp} we develop formal asymptotic
expansions of the eigenmodes and the associated eigenvalues of the magnetic
induction operator in power series in $\varepsilon^{1/2}$, where
$\varepsilon>0$ is the scale ratio. Finally, in section
\ref{cnc} we draw conclusions of the present study.

\section{Growth rates of large-scale magnetic modes}\label{num}

We explore numerically dynamos employing the \hbox{$\alpha$-effect}
or negative eddy diffusivity for generation of large-scale harmonically
amplitude-modulated magnetic modes
$${\bf b}=\e^{\I\varepsilon\bf q\cdot x}{\bf B(x},\varepsilon),$$
where the scale ratio $\varepsilon>0$ is small, $|{\bf q}|=1$ and
$\bf B(x,\varepsilon)$ has the same space periodicity as the flow (their growth
rates were calculated in \cite{RCZ}, see also \cite{VZ} for a detailed
presentation). The eigenvalue of the magnetic induction operator
$$\L:{\bf b}\mapsto\eta\nabla^2{\bf b}+\nabla\times({\bf v}\times{\bf b})$$
associated with the large-scale mode is denoted by $\lambda(\bf q)$.
Here $\eta$ is the magnetic molecular diffusivity and Re\,$\lambda$ the growth
rate of the magnetic mode $\bf b$ (a negative growth rate actually indicates
that the associated mode is decaying). The mode is solenoidal,
\BE\nabla\cdot{\bf b}=0,\EE{bs}
and the fluid is supposed to be incompressible, $\nabla\cdot{\bf v}=0$.

We now briefly review the relevant results of the multiscale analysis,
derived by expanding a large-scale mode and the associated eigenvalue
in power series in $\varepsilon$,
\se{bo}\be{\bf b}&=\sum_{n=0}^\infty{\bf b}_n({\bf X},{\bf x})\,\varepsilon^n,\label{beo}\\
\lambda&=\sum_{n=0}^\infty\lambda_n\varepsilon^n.\label{leo}\end{align}\end{subequations}

The $\alpha$-effect tensor $\A$ is a $3\times3$ matrix, whose $k$th column is
$\A_k=\LA{\bf v}\times{\bf S}_k\RA$, where ${\bf S}_k$ are $2\pi$-periodic
zero-mean solutions to three auxiliary problems
\be\L{\bf S}_k=&-{\partial{\bf v}\over\partial x_k}\qquad
\Leftrightarrow\qquad\L({\bf S}_k+{\bf e}_k)=0,\label{Seq}\\
\LA{\bf f}\RA=&(2\pi)^{-3}\int_{\T^3}{\bf f}({\bf X},{\bf x})\,\d{\bf x}
=\sum_{k=1}^3\LA{\bf f}\RA_k{\bf e}_k\nonumber\end{align}
denotes the spatial mean over the fast variables $\bf x$ (i.e., over
the periodicity cell $\T^3$), and ${\bf e}_k$ are unit vectors of the Cartesian
coordinate system.

In the presence of the $\alpha$-effect, the leading order term in the expansion
\rf{leo} of the eigenvalue $\lambda(\bf q)$ is $\varepsilon\lambda_1(\bf q)$;
the respective maximum large-scale growth rate in the slow time
$T_1=\varepsilon t$ is
\BE\gamma_\alpha=
\sqrt{\max(\alpha_1\alpha_2,\,\alpha_2\alpha_3,\,\alpha_1\alpha_3)},\EE{mgr}
where $\alpha_i$ are eigenvalues of the symmetrized tensor $\sA$,
$$\sAe_i^j=(\Ae_i^j+\Ae_j^i)/2.$$

For parity-invariant flows (i.e., $\bf v(x)\!=\!-v(-x)$), $\A=0$ implying
$\gamma_\alpha=0$ and
$\lambda({\bf q})=\varepsilon^2\lambda_2({\bf q})+{\rm O}(\varepsilon^3)$.
The large-scale generating mechanism is eddy diffusivity. The eddy diffusivity
tensor
\BE\De^l_{mk}=\LA{\bf Z}_l\cdot\left(2\eta{\partial{\bf S}_k\over\partial x_m}
+{\bf e}_m\times({\bf v}\times({\bf S}_k+{\bf e}_k))\right)\RA\EE{Dlmk}
involves three zero-mean solutions ${\bf Z}_l$ to auxiliary problems
$\L^*{\bf Z}_l={\bf v}\times{\bf e}_l$ for the adjoint operator
\BE\L^*:{\bf b}\mapsto\eta\nabla^2{\bf b}-{\bf v}\times(\nabla\times{\bf b}).\EE{La}
It is easy to see that
\BE{\bf Z}_k=\eta^{-1}\nabla^{-2}({\bf v}\times({\bf S}^-_k+{\bf e}_k)),\EE{Zl}
where ${\bf S}^-_k$ denotes
the solution to the problem \rf{Seq} stated for the reverse flow $-\bf v$.
The large-scale growth-rate in the slow time $T_2=\varepsilon^2t$ is
Re$\,\lambda_{2_\pm}({\bf q})+{\rm O}(\varepsilon)$, where
\se{dei}\be&\lambda_{2_\pm}({\bf q})=-\eta-{1\over2}\sum_{j,l,n}
(D^l_n-D^n_l)q_j\pm\sqrt d,\label{d2}\\
&d=\!\!\sum_{j,l,n}\!\left(\!((\sD^l_n)^2\!-\sD^l_l\,\sD^n_n)q_j^2\!-2q_jq_n(\sD^l_n\,
\sD^l_j\!-\sD^l_l\,\sD^n_j)\!\right)\!,\label{d3}\end{align}
both sums are over even permutations of indices 1, 2 and~3 (i.e.,
$\epsilon_{jln}=1$, where $\epsilon_{jln}$ is the unit antisymmetric tensor) and
\BE D^l_n=\sum_m\De^l_{mn}q_m,\qquad\sD^l_n=(D^l_n+D^n_l)/2.\EE{d1}\end{subequations}
The minimum eddy diffusivity is defined as
\BE\eta_{\rm ed}=\min_{|{\bf q}|=1}(-{\rm Re}\,\lambda_{2_\pm}({\bf q})).\EE{miE}

\begin{figure}[t]
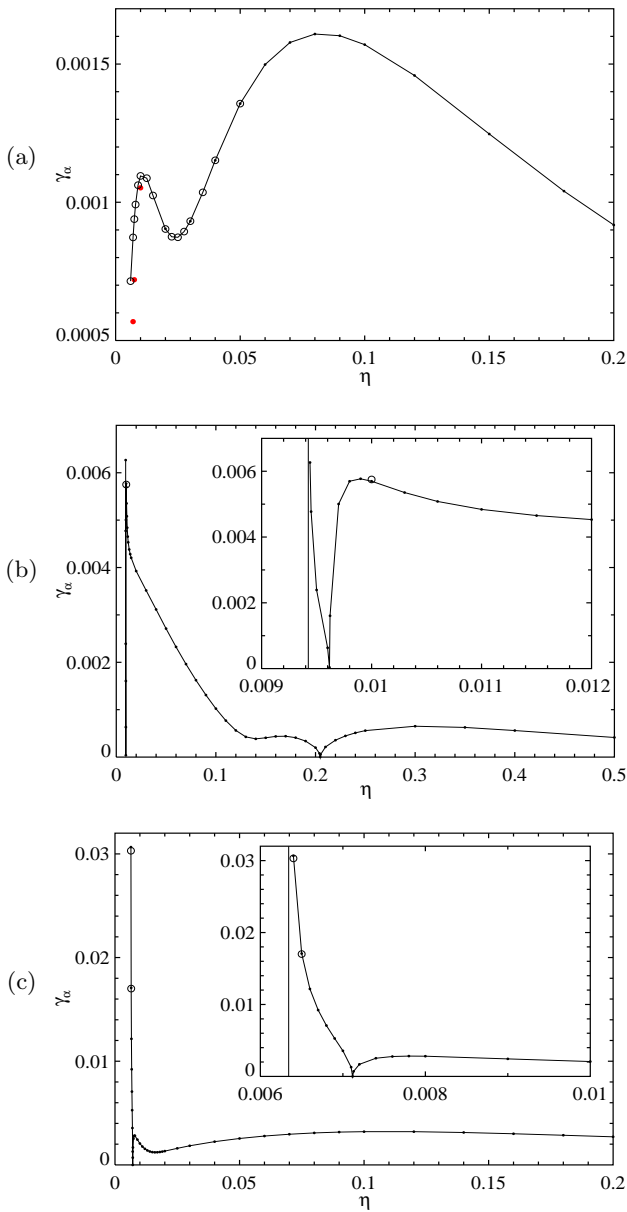

\centerline{\raisebox{3cm}{(a)} \ \includegraphics[width=3in,height=2in]{V1al.ps}}

~

\centerline{\raisebox{3cm}{(b)} \ \includegraphics[width=3in,height=2in]{V2al.ps}}

~

\centerline{\raisebox{3cm}{(c)} \ \includegraphics[width=3in,height=2in]{Lal.ps}}

\caption{Maximum slow-time growth rates \rf{mgr} of large-scale modes generated
by the $\alpha$-effect in sample flows \rf{vw} of zero velocity helicity
for $\bf w$ from
families V$_1$ (a), V$_2$ (b) and L (c) as functions of the molecular
diffusivity $\eta$. Insets in panels (b) and (c) show zooms of the graphs near
the critical diffusivity (indicated by thin vertical lines) for the onset
of the small-scale dynamo action. The resolution of computations is $128^3$
(solid dots) or $256^3$ (hollow circles) Fourier harmonics. Large solid dots
(red in the electronic version of the paper) in (a) show computations
in which the $128^3$ harmonics resolution is insufficient.}
\label{VVL}\end{figure}

\begin{figure}
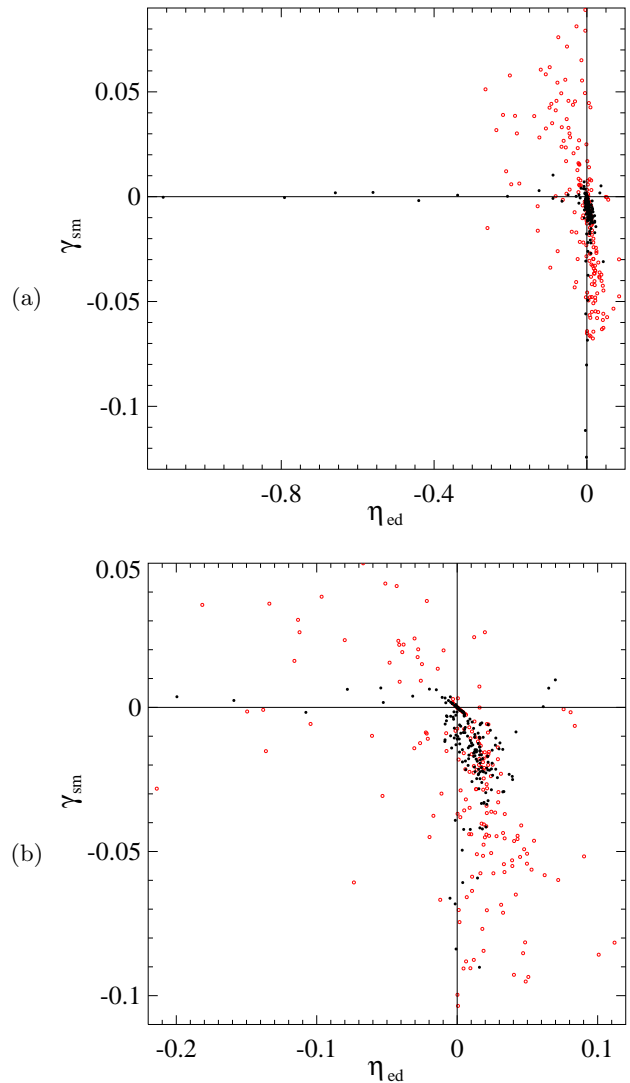

\centerline{\raisebox{3cm}{(a)} \ \includegraphics[height=2.75in]{flh01.ps}}

~

\centerline{\raisebox{3cm}{(b)} \ \includegraphics[height=2.75in]{flh02.ps}}
\caption{Dominant growth rates $\gamma_{\rm sm}$ of small-scale modes versus
minimum eddy diffusivity \rf{edm} in the primary cosine flows \rf{cos} (filled
circles) and \rf{wc} (hollow circles, red in the electronic version
of the paper) for $\eta=0.01$ (a) and 0.02 (b).}
\label{ras}\end{figure}

The fields ${\bf S}_k$ \rf{Seq} and the dominant small-scale
magnetic modes and their growth rates have been computed by the code \cite{Zh}
implementing the standard pseudo-spectral method. Typically, we have used
the resolution of $128^3$ Fourier harmonics, however, for the smallest magnetic
molecular diffusivities considered here the double resolution computations
with $256^3$ harmonics have been performed. The energy spectrum
of all fields used to construct graphs in Figs.~\ref{VVL} and~\ref{ras}
decays by at least 4 orders of magnitude.

\section{$\alpha$-effect in flows of zero helicity}\label{al}

\begin{figure}[!]
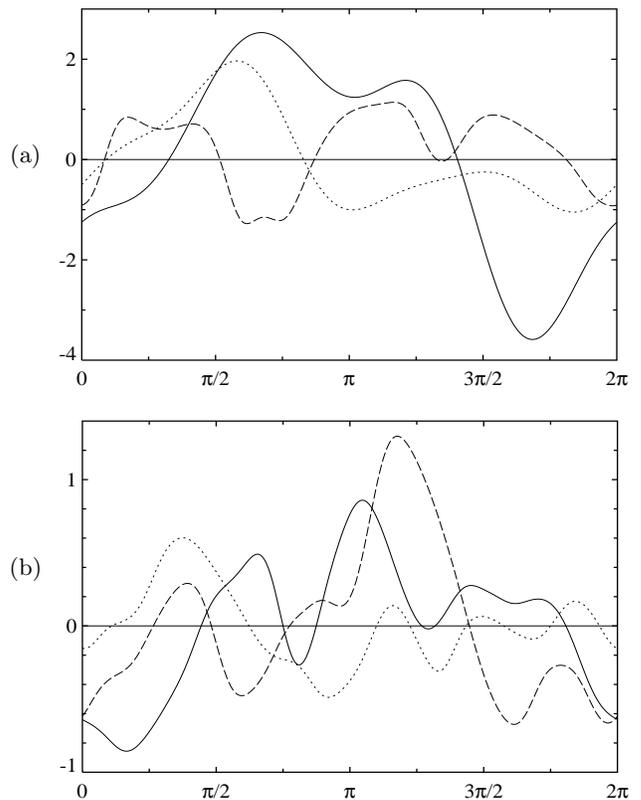

\centerline{\raisebox{3cm}{(a)} \ \includegraphics[width=3in,height=2in]{V1Ui.ps}}

~

\centerline{\raisebox{3cm}{(b)} \ \includegraphics[width=3in,height=2in]{V2Ui.ps}}
\caption{Graphs of $U_1$ (solid lines), $U_2$ (dashed lines) and $U_3$ (dotted
lines, vertical axes) as functions of the respective Cartesian coordinate
variable $x_i$ (horizontal axis) used to construct the zero velocity helicity
sample flows \rf{V1} and \rf{V2} for V$_1$ (a) and V$_2$ (b) family potentials
$\bf w$ (see \rf{vw}), \rf{W1} and \rf{W2}, respectively.}
\label{UU}\end{figure}

We have computed the maximum growth rates $\gamma_\alpha$ \rf{mgr}
of large-scale modes generated by the action of the \hbox{$\alpha$-effect} in
three sample flows \rf{vw},
\se{vf}\be{\bf v}^{{\rm V}_1}({\bf x})&=\beta^{{\rm V}_1}
(\dot U_1(C_3U_3\ddot U_2-C_2U_2\ddot U_3),\label{V1}\\
&\dot U_2(C_1U_1\ddot U_3
-C_3U_3\ddot U_1),\ \dot U_3(C_2U_2\ddot U_1-C_1U_1\ddot U_2),\nonumber\\
{\bf v}^{{\rm V}_2}({\bf x})&=\beta^{{\rm V}_2}
(U_1(C_3\dot U_2-C_2\dot U_3),\label{V2}\\
&U_2(C_1\dot U_3-C_3\dot U_1),\ U_3(C_2\dot U_1-C_1\dot U_2)),\nonumber\\
{\bf v}^{\rm L}({\bf x})&=\beta^{\rm L}
\nabla A\times\nabla B\label{L},\end{align}\end{subequations}
(the coefficients $\beta>0$ normalize the flows so that the r.m.s.~flow velocity
is unity) for vector potentials from families V$_1$, V$_2$ and L \cite{RCZ},
\pagebreak
\se{wf}\be{\bf w}^{{\rm V}_1}({\bf x})&=(C_1U_1\dot U_2\dot U_3,\ C_2\dot U_1U_2\dot U_3,\ C_3\dot U_1\dot U_2U_3),\label{W1}\\
{\bf w}^{{\rm V}_2}({\bf x})&=(C_1U_2U_3,\ C_2U_1U_3,\ C_3U_1U_2),\label{W2}\\
{\bf w}^{\rm L}({\bf x})&=(A\nabla B-B\nabla A)/2,\label{WL}\end{align}\end{subequations}
respectively. Here $C_i$ are arbitrary constants (satisfying $C_1+C_2+C_3=0$
in \rf{V1} and \rf{W1}); $U_i$ are arbitrary smooth $2\pi$-periodic functions
of $x_i$ (at least two of which are zero-mean in \rf{V2}); $A$ and $B$ are
eigenfunctions of the Laplace operator associated with the same eigenvalue;
dots denote differentiation of $U_i$ in $x_i$. We have employed the family
V$_1$ flow considered in \cite{RCZ} (see (62) in that paper)
to construct a sample flow \rf{V1} (the constitutive functions $U_i(x_i)$ are
shown in \xf{UU}(a)). In our sample flow \rf{V2}, $U_i$ are
random-coefficient sums of 20 Fourier harmonics whose energy spectra decay
by 3 orders of magnitude (see \xf{UU}(b)). In \rf{L}, the functions $A$ and $B$
are random-coefficient linear combinations of 72 Fourier harmonics that are
eigenfunctions of the Laplacian associated with the eigenvalue $-26$. Their
wave vectors are ($\pm3,\pm4,\pm1$) and ($\pm5,0,\pm1$), and those obtained
from these vectors by component permutations; the value $-26$ has been chosen
as the smallest one for which there exist two such triads of wave numbers.

For the same flows and molecular magnetic diffusivities $\eta$, we have computed
the maximum growth rates $\gamma_{\rm sm}$ of zero-mean small-scale
magnetic modes (i.e., eigenfunctions of the magnetic induction operator $\L$
that have the same spatial periodicity as the flow) and checked that
no small-scale dynamos operate simultaneously with our $\alpha$-effect dynamos.

The behavior of $\gamma_\alpha$ on varying molecular diffusivity $\eta$
in the three flows is drastically different. While for flows ${\bf v}^{{\rm V}_2}$
(see \xf{VVL}(b)) and ${\bf v}^{\rm L}$ (\xf{VVL}(c)) the maximum
slow-time growth rates increase to infinity when $\eta$ approaches from above
the critical diffusivity $\eta=\eta^{\rm cr}$ for the onset of the small-scale
field generation, for ${\bf v}^{{\rm V}_1}$ (\xf{VVL}(a)) it decays
on decreasing~$\eta$ (although this trend cannot be guaranteed to persist
for $\eta$ smaller than those shown). By \rf{mgr},
$\gamma_\alpha=0$ when the intermediate eigenvalue $\alpha_2$
of the symmetrized tensor $\sA$ vanishes; the graphs feature cusps of the form
$(\eta-\eta^{\rm cr})^{1/2}$ around such points. This occurs
for ${\bf v}^{\rm L}$ once, for ${\bf v}^{{\rm V}_2}$ twice, but never
for ${\bf v}^{{\rm V}_1}$. Such peculiarities in the behavior
of the maximum growth rates were also observed for flows of zero vorticity
helicity in \cite{RCZ}. \xf{VVL}(c) attests that $\gamma_\alpha$ can change
significantly under a tiny variation of~$\eta$.

\section{Singularities of the $\alpha$-effect\\growth rates at the onset\\
of the small-scale dynamo}\label{exp}

Our computations show that there exists a critical molecular diffusivity
$\eta=\eta^{\rm cr}>0$ for the onset of the small-scale dynamo. For this $\eta$,
a small-scale zero-mean mode ${\bf S}_0$ emerges in the kernel of $\L$:
$$\L{\bf S}_0=0,\qquad\LA{\bf S}_0\RA=0.$$
We observe that for the critical diffusivity, the maximum growth rate of large-scale
magnetic modes generated by the $\alpha$-effect has a singularity. As often
in physics, this indicates that near this diffusivity the original expansions
\rf{bo} of the mode and the associated eigenvalue break down, and we wiil
now investigate this. At first sight,
emergence of the new neutral short-scale mode requires introducing a new
amplitude for this mode and increasing the number of solvability conditions.
However, it is also necessary to take into account the possibility of emergence
of Jordan normal form cell associated with the eigenvalue zero
in the short-scale magnetic induction operator. Apart from these issues, the general
course of actions remains the same as in the general problem \cite{VZ}:
The size of the Jordan form cell implies the form of the power series in which
the large-scale mode and the associated eigenvalue of the large-scale magnetic
induction operator are expanded. From the eigenvalue equation expanded
in the power series, we deduce a hierarchy of equations for successive terms
in the series for the mode and the associated eigenvalue. These elliptic
equations in the fast variables are considered in consecutive order. First,
the solvability conditions are verified; this yields differential equations
in the slow variables for successive terms in the expansion of the mean field.
Second, we construct solutions to the equations in the short-scale variables
in terms of solutions to short-scale auxiliary
problems. In principle, all terms in the expansions can be found this way, but
we stop upon deriving a closed eigenvalue equation for the leading-order terms
in the expansions of the mean magnetic mode and the associated eigenvalue.

Any eigenfunction $\bf b$ of $\L$ (i.e., $\L{\bf b}=\lambda\bf b$), as well as
any generalized one (such that $\L^m{\bf b}=\lambda\bf b$ for an integer $m>0$)
has a non-zero mean only for $\lambda=0$. However, space-periodic
eigenfunctions (including generalized ones) of any elliptic operator constitute
a basis in the Lebesgue space of vector fields of the same spatial periodicity.
Thus, for $\eta=\eta^{\rm cr}$ the (possibly generalized) kernel of $\L$
involves at least 3 other (possibly generalized) eigenfunctions
${\bf S}_k\ (k=1,2,3)$ such that $\LA{\bf S}_k\RA$ constitute a basis in $\R^3$.

Evidently, the kernel of the adjoint operator \rf{La}
involves 3 constant vector fields ${\bf S}^*_k={\bf e}_k,\ k=1,2,3$.
The dimensions of the generalized kernels of $\L$ and its adjoint coincide,
as well as the Jordan normal form structures of the two operators. Therefore,
for $\eta=\eta^{\rm cr}$ there exists a zero-mean possibly generalized
eigenfunction ${\bf S}^*_0$ in the kernel of $\L^*$, such that
$\L^*{\bf S}^*_0=\bf Q$, where $\bf Q$ is a constant vector. Two possibilities
arise: ${\bf Q}=0$, any eigenfunction of $\L$ or $\L^*$
constituting an invariant subspace of the respective operator (there are no
generalized eigenfunctions in the two kernels), or ${\bf Q}\ne0$, in which case
each operator has two size 1 Jordan form cells and one size 2 cell, associated
with the eigenvalue~0. Computations confirm that the latter possibility
realizes and that the dimension of the generalized kernel of $\L$ generically
is four:
\se{sk}\be&\L{\bf S}_k=0,\ k=0,1,2;\quad\L{\bf S}_3={\bf S}_0;\label{s1}\\
&\LA{\bf S}_0\RA=0,\quad
\LA{\bf S}_k\RA\ (k=1,2,3)\ \mbox{constitute a basis in }\R^3;\label{s2}\\
&\L^*{\bf S}^*_k=0,\quad{\bf S}^*_k={\bf e}_k\ (k=1,2,3);\label{s3}\\
&\L^*{\bf S}^*_0={\bf Q}\ne0,\quad\LA{\bf S}^*_0\RA=0.\label{s4}
\end{align}\end{subequations}
We proceed to explore this case.

We assume henceforth in this section $\eta=\eta^{\rm cr}$ and make
no assumptions concerning the velocity (or any other) helicity.
The results that we obtain here are applicable whenever the generalized kernel
of the small-scale magnetic induction operator $\L$ is four-dimensional, and
the operator involves a $2\times2$ Jordan normal form cell associated
with the eigenvalue 0, whereby there exist the fields ${\bf S}_k$ and
${\bf S}^*_k$ with the properties \rf{sk}.

The solvability condition for a problem $\L\bf b=f$ is the orthogonality
of the r.h.s.~to the (non-generalized) kernel of the adjoint operator
in the functional Lebesgue space, which amounts to $\LA{\bf f}\RA=0$ (this
follows from the Fredholm alternative theorem for linear problems with compact
operators).

In contrast with \rf{bo}, the presence of the Jordan cell requires considering
expansions of a large-scale mode and the associated eigenvalue in power series
in $\varepsilon^{1/2}$ \cite{Ka,V87}:
\se{ble}\be{\bf b}&=\sum_{n=0}^\infty\wb_n({\bf X},{\bf x})\,\varepsilon^{n/2},\label{bex}\\
\lambda&=\sum_{n=0}^\infty\wl_n\varepsilon^{n/2}.\label{lex}\end{align}\end{subequations}
Substituting them into the eigenvalue equation $\L{\bf b}=\lambda\bf b$
yields a hierarchy of equations at successive orders $\varepsilon^n$:
\be\L\wb_n&+2\eta(\nabla\cdot\nabla_{\bf X})\wb_{n-2}
+\nabla_{\bf X}\times({\bf v}\times\wb_{n-2})
+\eta\nabla^2_{\bf X}\wb_{n-4}\nonumber\\
&=\sum_{m=0}^n\wl_{n-m}\wb_m.\label{hiN}\end{align}
Here the subscript $\bf X$ marks differential operators in slow variables;
the magnetic induction operator $\L$ is henceforth assumed
to involve differentiation in fast variables $\bf x$ only.
The solenoidality condition \rf{bs} at order $\varepsilon^{n/2}$ yields
an equation, whose mean and fluctuating parts are
\se{so}\begin{align}
\nabla_{\bf X}\cdot\langle\wb_n\rangle&=0,\label{mh}\\
\nabla_{\bf x}\cdot\wb_n+\nabla_{\bf X}\cdot\{\wb_{n-2}\}&=0.
\label{fh}\end{align}\end{subequations}

{\it Order $\varepsilon^0$ equation.}
Under our assumptions, the relevant solution to the first equation
in the hierarchy \rf{hiN},
$$\L\wb_0=\wl_0\wb_0,$$
is
$$\wb_0=\sum_{k=0}^2c_{0k}({\bf X}){\bf S}_k({\bf x}),\qquad\wl_0=0.$$

{\it Order $\varepsilon^{1/2}$ equation.}
The solvability condition for the second equation,
$$\L\wb_1=\wl_1\wb_0,$$
states $\langle\wb_0\rangle=0\ \Rightarrow\ c_{01}=c_{02}=0$ (since
$\LA{\bf S}_1\RA$ and $\LA{\bf S}_2\RA$ are linearly independent). Thus,
$$\wb_1=\sum_{k=0}^3c_{1k}({\bf X}){\bf S}_k({\bf x}),$$
where
\BE c_{13}=\wl_1c_{00}.\EE{c00}
A natural condition $\wb_0\ne 0$ (of the normalization sense) implies
$\wl_1\ne0$.

{\it Order $\varepsilon$ equation} is
\BE\L\wb_2+2\eta(\nabla\cdot\nabla_{\bf X})\wb_0
+\nabla_{\bf X}\times({\bf v}\times\wb_0)=\wl_2\wb_0+\wl_1\wb_1.\EE{f2}
In view of \rf{c00}, its solvability condition,
$$\nabla_{\bf X}\times\langle{\bf v}\times\wb_0\rangle=\wl_1\langle\wb_1\rangle,$$
is an eigenvalue-like problem for $\wl_1$ and $c_{13}$:
\BE\nabla_{\bf X}\times(\A c_{13})=\wl_1^2\sum_{k=1}^3\LA{\bf S}_k\RA c_{1k}.\EE{eid}
Here $\A$ denotes the tensor (actually, now a vector) of the magnetic
$\alpha$-effect
$$\A=\LA{\bf v}\times{\bf S}_0\RA.$$
This is consistent with Parker's \cite{P55} idea that
the interaction of fine structures of a flow (here,~$\bf v$) and magnetic
field (predominantly $c_{00}{\bf S}_0$) gives rise to a mean
e.m.f.~($c_{00}\A$) that may have a component, parallel to the large-scale
mean magnetic field (predominantly $\langle\wb_1\rangle$).

\begin{figure}[t]
\centerline{\includegraphics[width=3in]{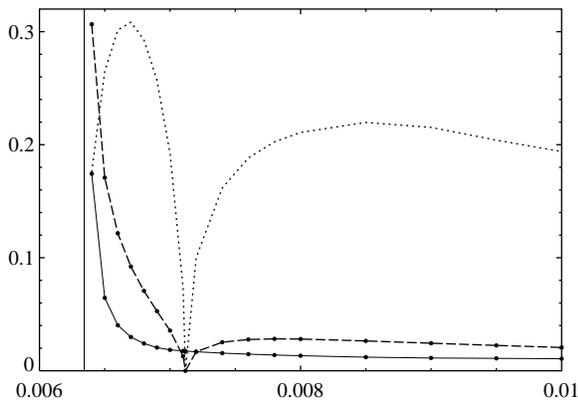}}
\caption{Graphs of $10\max_{|{\bf q}|=1}{\rm Re}\,\lambda_1({\bf q})$
(dashed line), $\max_{|{\bf q}|=1}{\rm Im}\,\lambda_1({\bf q})$ (solid line)
and their ratio (dotted line) as functions of the molecular eddy
diffusivity $\eta$ (horizontal axis). Small circles show the computed values.
Thin vertical line is the asymptote for the maximum growth rate
$\max_{|{\bf q}|=1}{\rm Re}\,\lambda_1({\bf q})$
located at the critical molecular diffusivity $\eta=\eta^{\rm cr}$,
for the onset of the small-scale dynamo.}\label{rir}\end{figure}

To solve the problem \rf{eid}, we assume that the large-scale
mode is amplitude-modulated by a Fourier harmonic,
$${\bf c}_1={\bf C}\e^{\I\bf q\cdot X},$$
\pagebreak
whereby
\BE\I{\bf q}\times\A C_3=\wl_1^2\sum_{k=1}^3\LA{\bf S}_k\RA C_k.\EE{ei2}
Here $\bf q$ and $\bf C$ are constant vectors, $|{\bf q}|=1$.
Scalar multiplying \rf{ei2} by $\LA{\bf S}_1\RA\times\LA{\bf S}_2\RA$ yields
\BE\wl_1({\bf q})=\pm(1+\I)\,
\sqrt{({\bf q}\times\A)\cdot(\LA{\bf S}_1\RA\times\LA{\bf S}_2\RA)
\over2(\LA{\bf S}_1\RA\times\LA{\bf S}_2\RA)\cdot\LA{\bf S}_3\RA}\EE{la1}
(by our assumptions, the denominator is non-zero). Evidently, there exists
a growing mode (Re\,$\wl_1>0$), unless the numerator vanishes; the maximum growth
rate is
$$\max_{|{\bf q}|=1}{\rm Re}\,\wl_1({\bf q})=
\sqrt{|\A\times(\LA{\bf S}_1\RA\times\LA{\bf S}_2\RA)|
\over2|(\LA{\bf S}_1\RA\times\LA{\bf S}_2\RA)\cdot\LA{\bf S}_3\RA|}.$$
Also, we determine from \rf{ei2}
$$c_{1k}=\epsilon_{k,3-k,3}\,{({\bf q}\times\A)\cdot(\LA{\bf S}_{3-k}\RA
\times\LA{\bf S}_3\RA)\over({\bf q}\times\A)\cdot(\LA{\bf S}_1\RA\times\LA{\bf S}_2\RA)}\,C_3$$
for $k=1,2$. The solenoidality of $\langle\wb_1\rangle$ \rf{mh} is equivalent to
${\bf q}\cdot\sum_{k=1}^3\LA{\bf S}_k\RA c_{1k}=0$, that follows
directly from \rf{ei2}.

We deduce from \rf{la1} that
$$\max_{|{\bf q}|=1}{\rm Im}\,\wl_1({\bf q})=
\max_{|{\bf q}|=1}{\rm Re}\,\wl_1({\bf q}).$$
This suggests that the ratio
of the maximum real and imaginary parts of the dominant term $\lambda_1$
in the expansion \rf{leo} of the eigenvalue for $\eta>\eta^{\rm cr}$
tends to unity when $\eta\to\eta^{\rm cr}$ from above. In order to test this
hypothesis, we plot in \xf{rir} the real and imaginary parts of $\lambda_1$
for the flow \rf{L} considered in section \ref{al} (the real part is multiplied
by 10 in order to unify the vertical scales of the three graphs), as well as
their ratio. Clearly, the ratio is far from the predicted limit value.
This is not very surprising: The singular behavior of Re$\,\lambda_1$
for $\eta\to\eta^{\rm cr}$ is offset by shrinking to zero of the radius
of convergence of the series \rf{leo}. Branches of eigenvalues of the magnetic
induction operator $\lambda(\eta,\varepsilon,\bf q)$ are continuous in the three
quantities on which they depend. However, both the real and imaginary parts
of $\lambda$ tend to 0 when $\varepsilon\to0$, and therefore there is no
continuity at $\varepsilon=0$ of the ratios Im$\,\lambda/{\rm Re}\,\lambda$ or
$\max_{|{\bf q}|=1}{\rm Im}\,\lambda/\max_{|{\bf q}|=1}{\rm Re}\,\lambda$.

Thus, from the equation for $n=2$ we obtain the leading-order term
$\wl_1\varepsilon^{1/2}$ in the expansion of the eigenvalue, and
large-scale amplitudes $c_{1k}({\bf X})$ for $k=1,2,3$ in the expansion of
the large-scale mode $\wb_1$. The leading-order term $\wb_1$ is now
completely determined by \rf{c00} (up to the coefficient $C_3$, which is
arbitrary because modes are defined up to a constant factor).
The solvability condition for \rf{f2} being satisfied, we find from this
equation the next term $\wb_2$ up to an arbitrary field from ker$\L$ and
the eigenvalue expansion term $\wl_2$. Following essentially the same
procedure, we can solve successively all equations \rf{hiN}, find all terms
in the expansions \rf{ble} and establish that solenoidality conditions \rf{so}
are satisfied.

\section{Negative eddy diffusivity\\in flows of zero velocity helicity}\label{ed}

We have computed the minimum magnetic eddy diffusivity \rf{miE} for flow \rf{vw}
\se{cos}\be v^{\rm C}_1=&\,\beta\left(({\bf a\cdot b})a_2+n^2b_2)\sin({\bf a\cdot x})\right.\label{c1}\\
&\,\left.+(({\bf a\cdot b})b_2+n^2a_2)\sin({\bf b\cdot x})\right)\sin nx_3,\nonumber\\
v^{\rm C}_2=&-\beta\left(({\bf a\cdot b})a_1+n^2b_1)\sin({\bf a\cdot x})\right.\label{c2}\\
&\,\left.+(({\bf a\cdot b})b_1+n^2a_1)\sin({\bf b\cdot x})\right)\sin nx_3,\nonumber\\
v^{\rm C}_3=&\,\beta n(a_1b_2-a_2b_1)(\cos({\bf a\cdot x})-\cos({\bf b\cdot x}))
\cos nx_3,\label{c3}\\
\beta=&\,2\left((n^4+({\bf a\cdot b})^2)(|{\bf a}|^2+|{\bf b}|^2)\right.\nonumber\\
&\,\left.+2n^2(({\bf a\cdot b})^2+|{\bf a}|^2|{\bf b}|^2)\right)^{-1/2}\label{c4}
\end{align}\end{subequations}
calculated for the family C ``cosine'' \cite{RCZ} potential
\be w^{\rm C}_1&=n(b_1\sin({\bf a\cdot x})+a_1\sin({\bf b\cdot x}))\cos nx_3,\nonumber\\
w^{\rm C}_2&=n(b_2\sin({\bf a\cdot x})+a_2\sin({\bf b\cdot x}))\cos nx_3,\label{wc}\\
w^{\rm C}_3&=-({\bf a\cdot b})(\cos({\bf a\cdot x})+\cos({\bf b\cdot x}))\sin nx_3,
\nonumber\end{align}
where ${\bf a}=(a_1,a_2,0)$ and ${\bf b}=(b_1,b_2,0)$ are constant horizontal
vectors. The flow \rf{cos} is parity-invariant relative points
Q$_j=(0,0,(j+1/2)\pi/n)$ for all integer $j$.

The symmetries of \rf{cos} and \rf{wc} are the same,
enabling us to apply the analysis \cite{RCZ} of the eddy diffusivity tensor
structure. Since translation by ${\bf a}=(\pi/n)\,{\bf e}_3$ reverses
the flow, ${\bf S}^-_n({\bf x})={\bf S}_n({\bf x+a})$ for all $n$, and solving
just the three auxiliary problems \rf{Seq} suffices for computing the eddy
diffusivity tensor $\D$ \rf{Dlmk}, \rf{Zl}. The translation antisymmetry
implies $\De^l_{mk}=-\De^k_{ml}$ for all $l,m,k$ and $\De^k_{mk}=0$.
Therefore, $\mD=0$, and for any wave vector
$\bf q$ eigenvalues \rf{dei} of the eddy diffusivity operator are real and
two-fold. Moreover, the flow \rf{cos} is symmetric in $x_3$ relative the points
Q$_j$, and therefore $\De^l_{mk}=0$ if all indices $l,m,k$ do not exceed 2,
or precisely two of them are equal to 3. Consequently,
$$\lambda_2({\bf q})=-\eta+\De^2_{31}q_3^2+\De^3_{12}q_1^2
+(\De^3_{22}+\De^1_{13})q_1q_2+\De^1_{23}q_2^2,$$
and the minimum eddy diffusivity \rf{miE} is
\be\eta_{\rm ed}=\eta&-\max\Big(\De^2_{31},\ {1\over2}\Big(\De^3_{12}+\De^1_{23}\nonumber\\
&+\sqrt{(\De^3_{12}-\De^1_{23})^2+(\De^3_{22}+\De^1_{13})^2}\,\Big)\!\Big).
\label{edm}\end{align}

Generation of small- and large-scale fields was investigated
in \cite{RCZ} for a set of 183 ``primary'' sample flows \rf{wc} satisfying
the following conditions: 1) the horizontal vectors ${\bf a}=(a_1,a_2,0)$ and
${\bf b}=(b_1,b_2,0)$ have integer components such that $|a_i|\le 3$,
$|b_i|\le 3$, and $n\le3$ is a positive integer; 2) the largest common divisor
of the four numbers $a_1,a_2,b_1,b_2$ is 1; 3) the flows are
non-planar (by the Zeldovich \cite{Zel} antidynamo theorem planar flows
are irrelevant as non-dynamos); 4) no primary flow
can be mapped into another one by reflections and their combinations. Applying
the same rules, here we have selected for numerical study a set of 199 primary
flows \rf{cos} (the number has increased, because in contrast to \rf{wc}, flows
\rf{cos} for orthogonal $\bf a$ and $\bf b$ are non-planar).

\begin{table}
\caption{Primary flows \rf{cos} and \rf{wc} as small- and large-scale dynamos.
Numbers of the primary cosine flows \rf{cos} and \rf{wc} (in parenthesis,
from \cite{RCZ}) falling into the specified categories are shown.}
\center\begin{tabular}{ccccc}\hline\hline
&\multicolumn{2}{c}{$\eta=0.01$}&\multicolumn{2}{c}{$\eta=0.02$}\\%\cline{2-5}
&$\eta_{\rm ed}<0$&$\eta_{\rm ed}>0$&$\eta_{\rm ed}<0$&$\eta_{\rm ed}>0$\\\hline
Small-scale dynamo&33 (61)&\phantom{13}3 (12)&28 (27)&\phantom{13}4\phantom{13} (4)\\%\hline
No small-scale dynamo&36 (25)&127 (85)&23 (20)&144 (132)\\\hline\hline
\end{tabular}\label{tab1}\end{table}

The distributions of the dominant fast-time growth rates of small-scale magnetic
modes and of the minimum eddy diffusivity computed for the primary flows
\rf{cos} for $\eta=0.01$ and 0.02 are shown in \xf{ras} (see also
Table~\ref{tab1}) in comparison with the similar distributions
for the normalized primary flows \rf{wc}. More than a half
of the primary flows \rf{cos} (for $\eta=0.01$, roughly 1.5 times more than
for \rf{wc}) are neither \hbox{small-,} nor large-scale dynamos. For both
sets, the second largest category are flows that can generate both small-
and large-scale fields. While a comparable number of flows of our prime
interest is found that are incapable of small-scale generation but feature
negative eddy diffusivity, only a few flows in which eddy diffusivity is
positive can generate small-scale fields. In \xf{ras}, the growth rates
concentrate significantly closer to the origin for flows \rf{cos} than
for \rf{wc}, although, unlike the flows \rf{wc}, some flows \rf{cos} feature
for $\eta=0.01$ a strong negative eddy diffusivity; the small respective growth
rates of small-scale fields suggest that for them $\eta=0.01$ is close
to the critical value for the onset of the small-scale generation.

\section{Concluding remarks}\label{cnc}

We have considered kinematic dynamos powered by steady flows, whose velocity
helicity density vanishes in every point
in space. The flows employed in simulations, \rf{vf} and \rf{cos}, have been
constructed as normalized curls \rf{vw} of space-periodic steady solenoidal
flows $\bf w$, whose vorticity helicity (in other words, kinetic helicity)
is pointwise zero; these $\bf w$ belong to four analytically defined families
V$_1$, V$_2$, L \rf{wf} and C \rf{wc} introduced in~\cite{RCZ}. All flows
studied here have an identically zero helicity spectrum, except for
${\bf v}^{\rm L}$ \rf{L} (see \xf{hs}). Generation of large-scale fields has
been investigated in the limit of high scale separation by computing
the magnetic $\alpha$-effect or eddy diffusivity tensors formerly derived
by applying the multiscale formalism (see, e.g., \cite{VZ}). We have established
that both mechanisms generate large-scale magnetic field and a significant part
of the employed flows generate small-scale field for the considered molecular
diffusivities (corresponding to moderate local magnetic Reynolds numbers below 200).
This is our main finding: kinematic generation of small- or large-scale
magnetic field does not require flows to have a non-zero velocity helicity or
helicity spectrum. This reinforces the skepticism of \cite{RCZ} concerning
the ability of any helicity-like quantity to characterize
reliably the flow efficiency as a dynamo.

We observe a fast increase, from 0 to $\infty$, of the maximum (over
the direction of the large-scale wave vector $\bf q$) large-scale
growth rate due to the action of the \hbox{$\alpha$-effect}, $\gamma_\alpha$, under
a tiny variation of molecular eddy diffusivity decreasing from $\eta=0.007118$
(for which the intermediate eigenvalue of the symmetrized $\alpha$-tensor
vanishes) to $\eta^{\rm cr}=0.006342$ (the critical diffusivity for the onset
of the small-scale dynamo action), see \xf{VVL}(c). Let us note that
(like in the case of infinitely negative eddy diffusivity), this does not imply
an unphysical singular behavior of the large-scale dynamo: When
the $\alpha$-effect tensor is non-zero (the generic case), it determines
the leading-order term in the expansion in the scale ratio $\varepsilon$
of an eigenvalue of the large-scale magnetic induction operator,
$\lambda=\sum_{n=0}^\infty\lambda_n\varepsilon^n$. This term is
$\varepsilon\lambda_1$, where $\lambda_1$ is an eigenvalue of the magnetic
$\alpha$-effect operator ${\bf b}\mapsto\nabla_{\bf X}(\A{\bf b})$. However,
the power series has a finite radius of convergence apparently tending
to zero for $\eta\to\eta^{\rm cr}$, so that $\lambda$ remains finite or even
tends to zero. As we have shown in section \ref{exp}, the relevant expansion
of $\lambda$ at $\eta=\eta^{\rm cr}$ is in the power series
in $\sqrt\varepsilon$. The expansion has revealed that this large-scale dynamo
has an unusual feature: the amplitude of the mean magnetic field is order
$\sqrt\varepsilon$ smaller than the amplitude of the fluctuating component
of the field.

The dynamos explored here are slow: Fluid particle
trajectories for flows \rf{V1} and \rf{V2} have first integrals
$\sum_{m=1}^3\int(C_mU_m/\dot U_m)\d x_m$ and
$\sum_{m=1}^3\int(C_m/U_m)\d x_m$, respectively, and for flow \rf{L}
two first integrals $A$ and~$B$. For a flow \rf{cos}, the trajectories
belong to vertical surfaces, whose intersections with horizontal planes satisfy
the differential equation which is the ratio of \rf{c1} and \rf{c2}.
This is incompatible with
a chaotic behavior of the trajectories required for fast dynamos \cite{V89,KY}.
An interesting mathematical problem is to construct a pointwise non-helical
steady flow acting as fast dynamo (or at least lacking global first integrals).

\begin{acknowledgments}
RC was supported by the project POCI-01-0145-FEDER-006933/SYSTEC financed
by ERDF through COMPETE 2020 and by FCT (Portugal). The authors would like
to thank the anonymous Referee, whose comments helped us to significantly
improve the paper.
\end{acknowledgments}

\end{document}